# ON AN OPTIMIZATION TECHNIQUE USING BINARY DECISION DIAGRAM

Debajit Sensarma[#1], Subhashis Banerjee[#1], Krishnendu Basuli[#1],Saptarshi Naskar[#2], Samar Sen Sarma[#3]

[#1]West Bengal State University, West Bengal, India

debajit.sensarma2008@gmail.com, mail.sb88@gmail.com, Krishnendu.basuli@gmail.com

[#2]Sarsuna College, West Bengal, India

sapgrin@gmail.com

[#3]University Of Calcutta, West Bengal, India

Sssarma2001@yahoo.com

## ABSTRACT

*Two-level logic minimization is a central problem in logic synthesis, and has applications in reliability analysis and automated reasoning. This paper represents a method of minimizing Boolean sum of products function with binary decision diagram and with disjoint sum of product minimization. Due to the symbolic representation of cubes for large problem instances, the method is orders of magnitude faster than previous enumerative techniques. But the quality of the approach largely depends on the variable ordering of the underlying BDD. The application of Binary Decision Diagrams (BDDs) as an efficient approach for the minimization of Disjoint Sums-of-Products (DSOPs). DSOPs are a starting point for several applications.*

*The use of BDDs has the advantage of an implicit representation of terms. Due to this scheme the algorithm is faster than techniques working on explicit representations and the application to large circuits that could not be handled so far becomes possible. Theoretical studies on the influence of the BDDs to the search space are carried out. In experiments the proposed technique is compared to others. The results with respect to the size of the resulting DSOP are as good or better as those of the other techniques.*

## INTRODUCTION

A Binary Decision Diagram is a rooted directed acyclic graph. It has one or two terminal nodes of out-degree zero labeled 0 or 1, and a set of variable nodes also called as branch node of out-degree two. Fig 1 depicts a BDD. The root node 'a' in Fig 1. Have two successors indicated by descending lines. One of the successors is drawn as a dashed line, called **'low'** and other is drawn as a solid line, called **'high'**. These branch nodes define a path in the diagram for any values of Boolean variables. The '0' and '1' nodes also called the sink node. If low branch is being followed from the root, then that path will reach to sink node '0' and if high branch is being followed, then the path will reach to sink node '1'. The BDD obeys two important restrictions. First, it must be ordered. Second, a BDD must be reduced, in the sense that it doesn't waste space. BDD's are well-known and widely used in logic synthesis and formal verification of integrated circuits. Due to the canonical representation of Boolean functions they are very suitable for formal verification problems and used in a lot of tools to date [**15, 16, 18**].

A DSOP is a representation of a Boolean function as a sum of disjoint cubes. DSOPs are used in several applications in the area of CAD, e.g. the calculation of spectra of Boolean functions or as a starting point for the minimization of Exclusive-Or-Sum-Of-Products (ESOPs).

A hybrid approach for the minimization of DSOPs relying on BDDs in combination with structural methods has recently been introduced in. It has been shown that BDDs are applicable to the problem of DSOP minimization [**6**].

Given a BDD of a Boolean function, the DSOP can easily be constructed: each one-path [**7**], i.e. a path from the root to the terminal 1 vertex, corresponds to a cube in the DSOP, and moreover, different one-paths lead to disjoint cubes. For the construction of the BDD the variables of the Boolean function are considered in a fixed order. The permutation of the variables largely influences the number of one-paths in the BDD and thus the number of cubes in the corresponding DSOP. Additionally, the importance of choosing a good variable order to get a small DSOP has theoretically been supported.

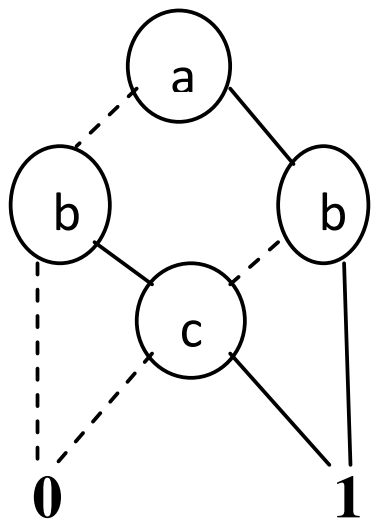

**Fig 1:** The Binary Decision Diagram

## 1.1 MOTIVATION

As minimizing Boolean functions with many variables is n NP-Complete problem, so the existence of a polynomial-time algorithm for minimizing Boolean circuits is unlikely. So, the motivation of this work is to minimize the Boolean sum of product function by finding the minimal irredundant expression. Here Binary Decision Diagram (BDD) is used for finding disjoint cubes first, because BDD is the compact representation of a Boolean function, but it highly depends on variable ordering. Then this disjoint cubes are minimizes to get the minimal expression. In Quine-McCluskey method, it can be shown that for a function of n variables the upper bound on the number of prime implicantes is $3^n/n$. In this project a heuristic algorithm is used to minimize the upper bound of the prime implicant generation and it gives the near optimal solution.

## 1.2 BINARY DECISION DIAGRAMS

A BDD is a directed acyclic graph $G_f = (V, E)$ that represents a Boolean function $f: B^n \rightarrow B^m$. The Shannon decomposition $g = x_i g_{xi} + x_{i'} g_{xi'}$ is carried out in each internal node v labeled with label (v) = $x_i$ of the graph, therefore v has the two successors then (v) and else (v). The leaves are labeled with 0 or 1 and correspond to the constant Boolean functions. The root node root ($G_f$ ) corresponds to the function f. In the following, BDD refers to a reduced ordered BDD (as defined in **[9]**) and the size of a BDD is given by the number of nodes.

**DEFINITION**

A one-path in a BDD $G_f = (V, E)$ is a path

$p = (v_0 ,\ldots, v_{l-1}, v_l)$;

$v_i$ € V; $(v_i, v_{i+1}) \in E$

with $v_0 = root(G_f)$ and $label(v_l) = 1$. p has length l + 1.

$P_1$ ($G_f$) denotes the number of all different one-paths in the BDD $G_f$.

### 1.3 BDD AND DSOP

Consider a BDD Gf representing the Boolean function $f(x_1,\ldots, x_n)$. A one path

$p = (v_0 ,\ldots, v_l)$ of length l + 1 in Gf corresponds to an (n - l)-dimensional cube that is a subset of $ON(f)^1$. The cube is described by:

$m_p$ =n $l_i$ for i=0…l-1; where

$l_i$ = label ($v_i$); if $v_i$+1 = else ($v_i$)

label ($v_i$); if $v_i$+1 = then ($v_i$)

Two paths p1 and p2 in a BDD are different if they differ in at least one edge. Since all paths originate from root ($G_f$), there is a node v where the paths separate. Let label (v) = $x_i$. Therefore one of the cubes includes xi, the other $x_i$. Hence, the cubes mp1 and mp2 are disjoint.

Now the DSOP can easily be built by summing up all cubes corresponding to the one-paths.

<u>Remark 1</u> : Let Gf be a BDD of $f(x_1 ,\ldots, x_n)$ and M1 be the set of one-paths

in $G_f$ . Then $G_f$ represents the DSOP

$$\sum m_p \text{ where } p \in M_1$$

where $m_p$ is the cube given above.

From this it is clear that the number of cubes in the DSOP represented by $G_f$ is equal to $P_1(G_f)$. Thus, as opposed to the usual goal of minimizing the number of nodes in a BDD, here the number of one-paths is minimized. Known techniques to minimize the number of nodes can be used to minimize the number of paths by changing the objective function. One such technique is sifting. A variable is chosen and moved to any position of the variable order based on exchange of adjacent variables. Then it is fixed at the best position (i.e. where the smallest BDD results), afterwards another variable is chosen. No variable is chosen twice during this process.

## 2. TERMS RELATED TO SOP MINIMIZATION

**Unate function:** A function that is monotonically increasing or decreasing in each of its variable is called unate function.
**binate function**: a function that is not unate. This can also be used to mean a cover of a function that is not unate.

**canonical cover/solution**: the SOP cover of a function that contains only minterms, and thus has not at all been reduced.

**cube**: a one-dimensional matrix in the form of an implicant. Two cubes are said to be **disjoint** if their intersection of the set of minterms is null. The intersection is the operation of conjunction (i.e. the Boolean AND operation).

**Espresso algorithm**: an algorithm that minimizes SOP functions.
**essential prime implicant**: a prime implicant that the cover of a function must contain in order to cover the function

**implicant**: an ANDed string of literals. It is a term in an SOP function.
**literal**: an instance of a boolean variable. It may be the variable complemented, uncomplemented, or ignored (don't-care). In matrix representations or the Q-M algorithm, it may have a value of 0, 1, or 2/X, corresponding to complemented, uncomplemented, and don't-care, respectively.

**matrix representation of a function or implicant**: The rows of a two-dimensional matrix representation of a function are the implicants of the function. The columns of a one-dimensional matrix representation of an implicant are the literals of the implicant.

**minterm**: an implicant that contains exactly one literal for each variable. It is not at all simplified.

**monotone decreasing**: A function is monotone decreasing in a variable if changing the value of the variable from 0 to 1 results in the output of the function being 0.

**monotone increasing**: A function is monotone increasing in a variable if changing the value of the variable from 0 to 1 results in the output of the function being 1.

**prime implicant**: an implicant that cannot be further reduced by adjacency

**Quine-McCluskey (Q-M) algorithms**: two algorithms that minimize a Boolean function. The first algorithm finds all prime implicants, and the second algorithm eliminates nonessential prime implicants.

## 3. PROPOSED ALGORITHM

In this method at first from the given truth table suitable variable order is chosen based on Shannon entropy measurement, then binary decision diagram is made considering this variable order. After that disjoint cubes are calculated by following the 1-path of the BDD. Then from that the covering matrix is created where columns represent the variables and row represents the applicants or disjoints cubes. Then selecting the most binate variables and by unite simplification the ultimate minimized sop function is obtained.

### 3.1 ALGORITHM

Step 1: Generation of truth table.
Step 2: Variable reordering using Shannon entropy measure-ment **[3, 9]** and create Binary Decision Diagram **[8, 14, 15]**.

Step 3: Finding Disjoint Cubes from Binary decision Diagram **[6]**.

Step 4: Disjoint cube minimization using Binate Covering with Recursion and Unate Simplification Method **[12,13]**.

### 3.2 EXPLANATION

**Step 1**:

The truth table is generated from given Boolean expression.

**Step2**:

Choosing right variable order is very important for constructing Binary Decision Diagram, because if bad variable order is chosen then number of 1-paths can be increased; even number of nodes in the BDD may be increased exponentially.

The measures of a variable's importance are based on information theoretic criteria, and require computation of entropy of a variable. Entropy measures can be quite effective in distinguishing the importance of variables. It is well known that a central problem in using OBDD is the severe memory requirements that result from extremely large OBDD size that arise in many instances.OBDD sizes are unfortunately very sensitive to the order chosen on input variables. Determining the optimal order is a co-NP complete problem **[9]**.Variable ordering heuristics can be classified as either static or dynamic approaches. A static approach , analyzes the given circuit/function and, based on its various properties, determines some variable order which has a high "Probability" of being effective. In dynamic approach to compute variable ordering, one starts with an initial order, which is analyzed and permuted at internal points in the circuit/function, such that some cost function is minimized.

**Step 3**:
In this step Disjoint Cubes from Binary decision Diagram are found by following the 1-path **[6]**.

**Step 4:**
Cover matrix is found from the resultant disjoint cubes and this is simplified using Unate Recursive Paradigm [**13**].

## 4. ILLUSTRATION WITH AN EXPLANATION

f(a,b,c,d)=∑(1,5,6,9,12,13,14,15)

Step1: *Given the truth table.*

| a | b | c | d | f |
|---|---|---|---|---|
| 0 | 0 | 0 | 0 | 0 |
| 0 | 0 | 0 | 1 | 1 |
| 0 | 0 | 1 | 0 | 0 |
| 0 | 0 | 1 | 1 | 0 |
| 0 | 1 | 0 | 0 | 0 |
| 0 | 1 | 0 | 1 | 1 |
| 0 | 1 | 1 | 0 | 1 |
| 0 | 1 | 1 | 1 | 0 |
| 1 | 0 | 0 | 0 | 0 |
| 1 | 0 | 0 | 1 | 1 |
| 1 | 0 | 1 | 0 | 0 |
| 1 | 0 | 1 | 1 | 0 |
| 1 | 1 | 0 | 0 | 1 |
| 1 | 1 | 0 | 1 | 1 |
| 1 | 1 | 1 | 0 | 1 |
| 1 | 1 | 1 | 1 | 1 |

With this variable order the Binary Decision Diagram is:

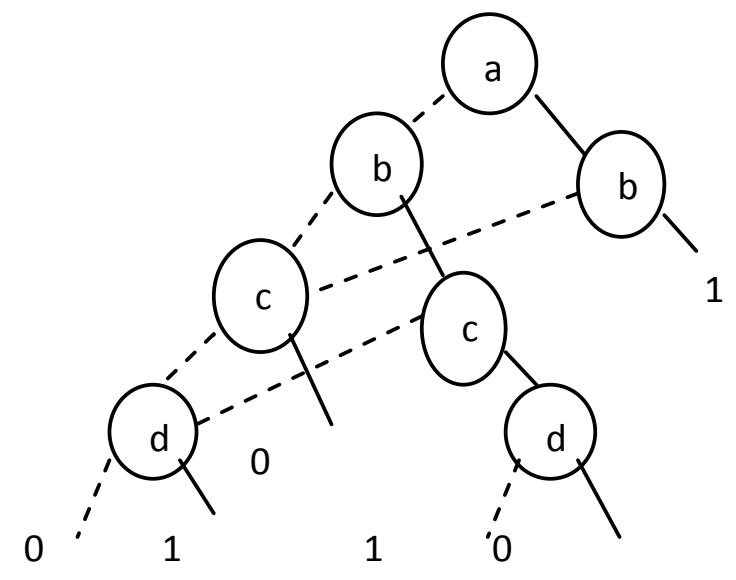

Number of nodes=7

<u>Step 2:</u> *Variable ordering by calculating Entropy and choosing most ambiguous variables.*

I(a,0)=0.954

I(a,1)=0.954

**E(a)=0.954**

I(b,0)=0.811

I(b,1)=0.811

**E(b)=0.811** Select 'b' as the first splitting variable.

I(c,0)=0.954

I(c,1)=0.954

**E(c)=0.954**

I(d,0)=0.954

I(d,1)=0.954

**E(d)=0.954**

For b=0 the truth table is:

| a | c | d | f |
|---|---|---|---|
| 0 | 0 | 0 | 0 |
| 0 | 0 | 1 | 1 |
| 0 | 1 | 0 | 0 |
| 0 | 1 | 1 | 0 |
| 1 | 0 | 0 | 0 |
| 1 | 0 | 1 | 1 |
| 1 | 1 | 0 | 0 |
| 1 | 1 | 1 | 0 |

I(a,0)=0.811

I(a,1)=0.811

**E(a)=0.811**

I(c,0)=1

I(c,1)=0

**E(c)=0.5**

I(d,0)=0

I(d,1)=1

**E(d)=0.5**

For b=1 the truth table is:

| a | c | d | f |
|---|---|---|---|
| 0 | 0 | 0 | 0 |
| 0 | 0 | 1 | 1 |
| 0 | 1 | 0 | 1 |
| 0 | 1 | 1 | 0 |
| 1 | 0 | 0 | 1 |
| 1 | 0 | 1 | 1 |
| 1 | 1 | 0 | 1 |
| 1 | 1 | 1 | 1 |

I(a,0)=1

I(a,1)=0

**E(a)=0.5** Select ‘a’ as the next splitting variable.

I(c,0)=0.811

I(c,1)=0.811

**E(c)=0.811**

I(d,0)=0.811

I(d,1)=0.811

**E(d)=0.811**

If we proceed like this we come up with the variable order →**b,a,c,d**

With this variable order the Binary Decision Diagram is:

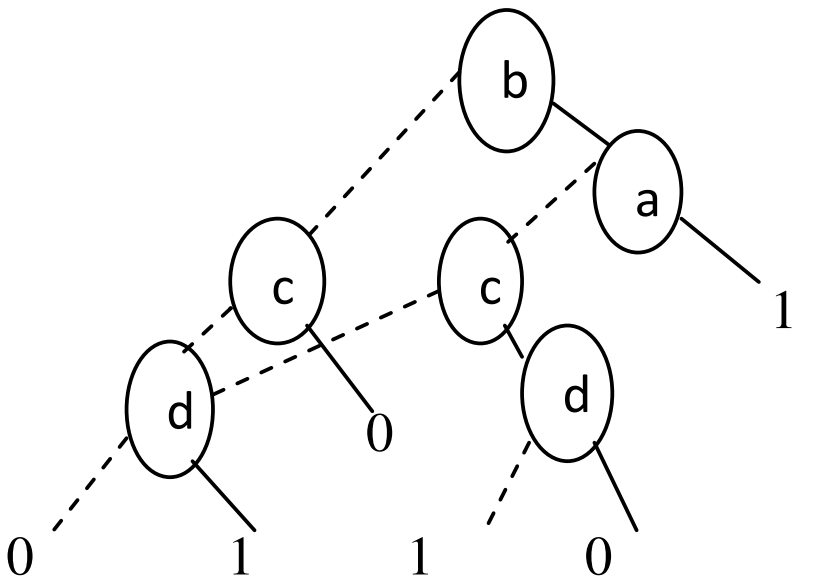

Number of nodes=6

<u>Step 3</u>: *Finding Disjoint Cubes from Above Binary Decision Diagram.*

The Disjoint Cubes are: **ab + a’bcd’+ b’c’d + a’bc’d**

<u>Step 4</u>: *Binate Covering with Recursion.*

The Covering Matrix is:

| | a | b | c | d |
|---|---|---|---|---|
| ab | 1 | 1 | 2 | 2 |
| a’bcd’ | 0 | 1 | 1 | 0 |
| b’c’d | 2 | 0 | 0 | 1 |
| abc’d | 0 | 1 | 0 | 1 |

<u>Case 1</u>: *Binate Select.*

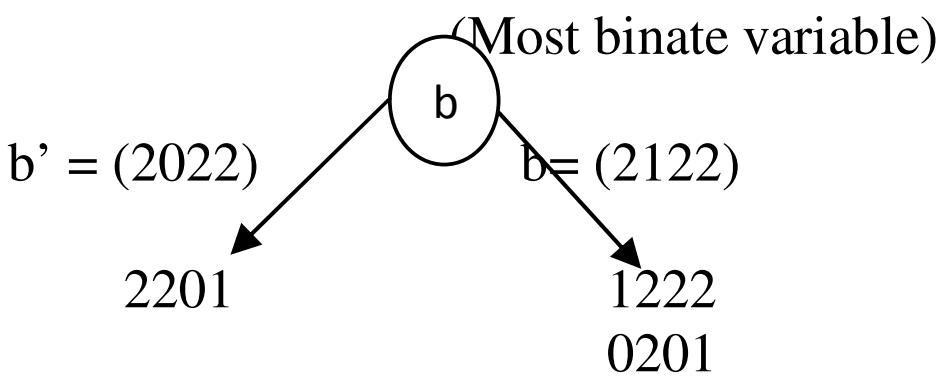

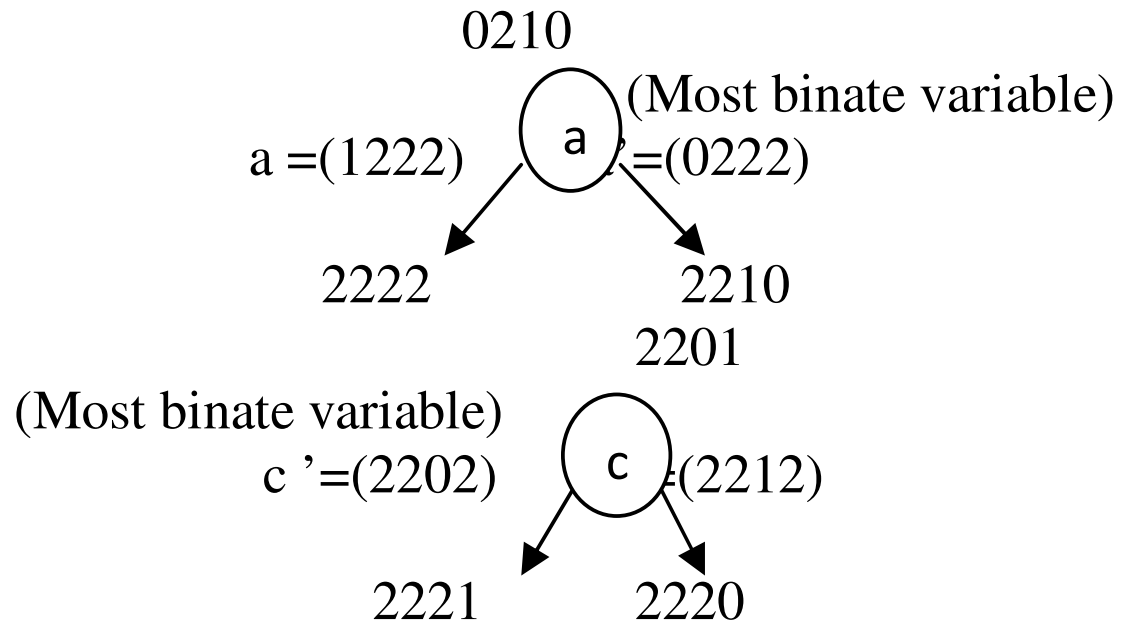

Case 2: *Merge.*

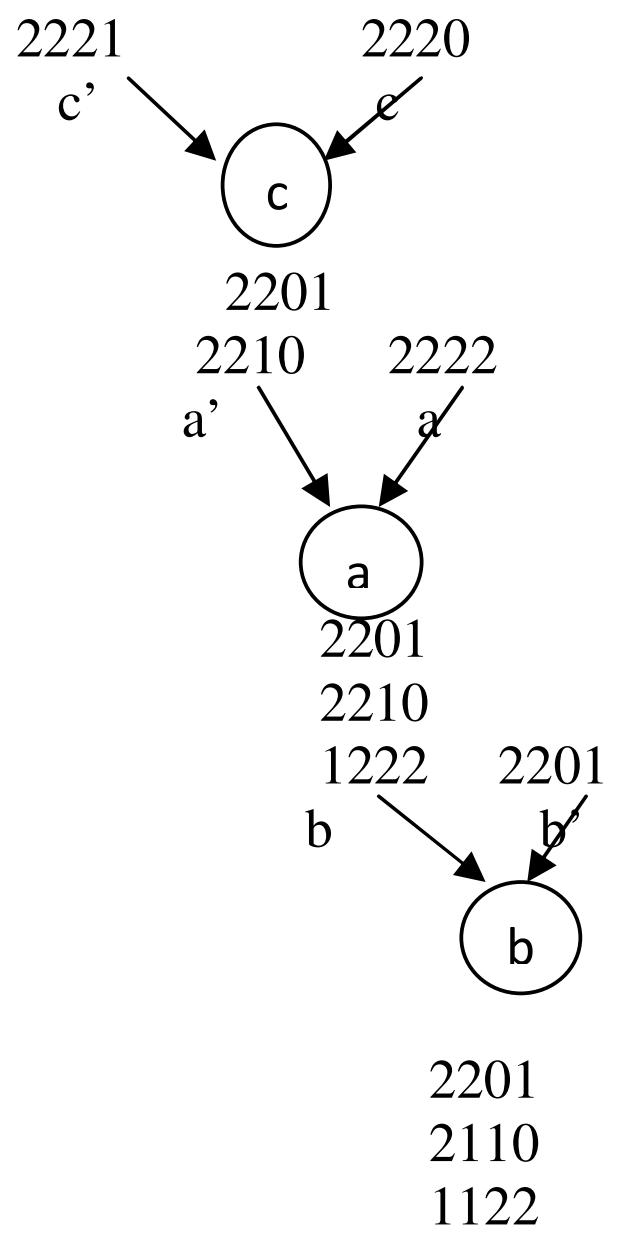

After simplification the expression is: **ab + c'd + bcd'**.

**KARNAUGH MAP REPRESENTATION**

f(a,b,c,d)=∑(1,5,6,9,12,13,14,15)

| cd / ab | 00 | 01 | 11 | 10 |
|---|---|---|---|---|
| 00 | | 1 | | |
| 01 | | 1 | | 1 |
| 11 | 1 | 1 | 1 | 1 |
| 10 | | 1 | | |

## 5. RESULT

This program is done on Intel Pentium 4 CPU, 2.80 GHz and 256 MB of RAM and with Visual c++ 6.0 standard edition. The result is presented here with the following figures.

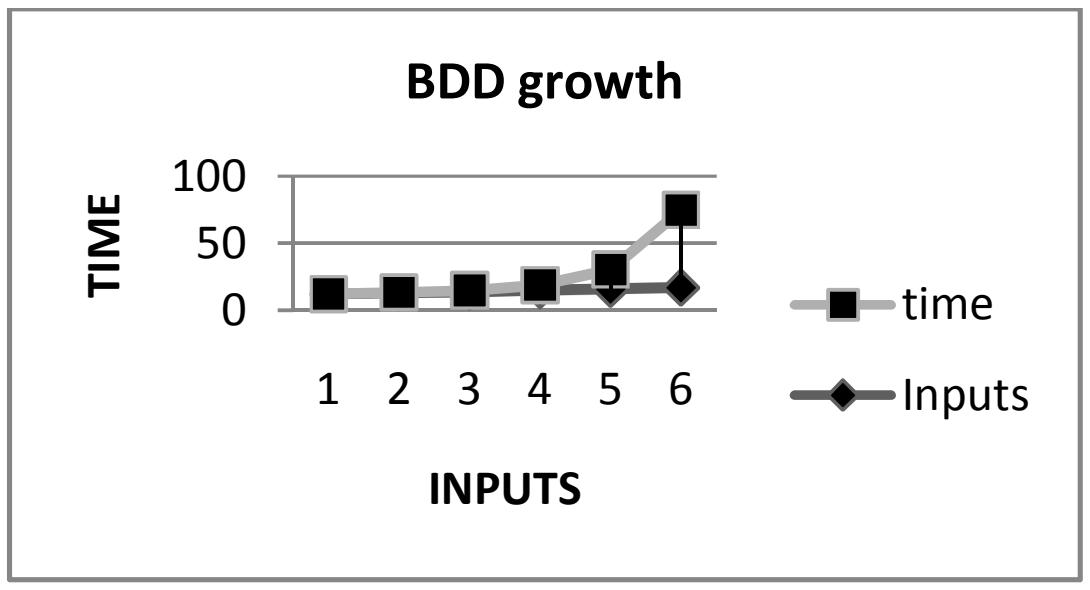

**Fig 2:** growth of Binary Decision Diagram. Here X-axis represents the number of inputs and Y-axis represents the required time.

Here growth of the creation of Binary decision diagram with proposed method is shown with respect to 6 variables.

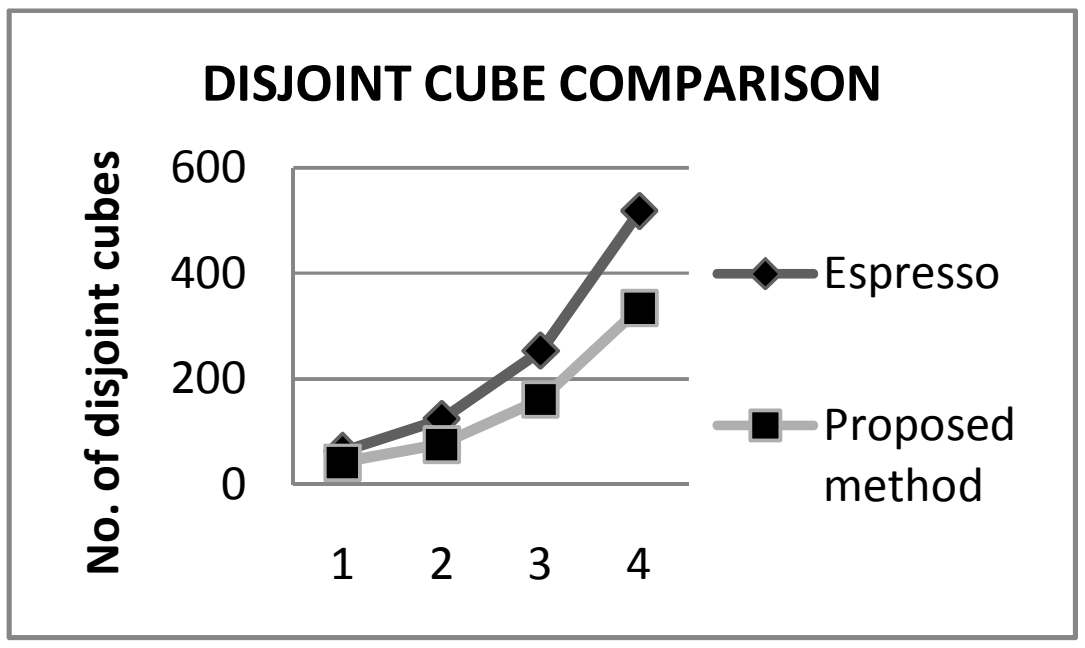

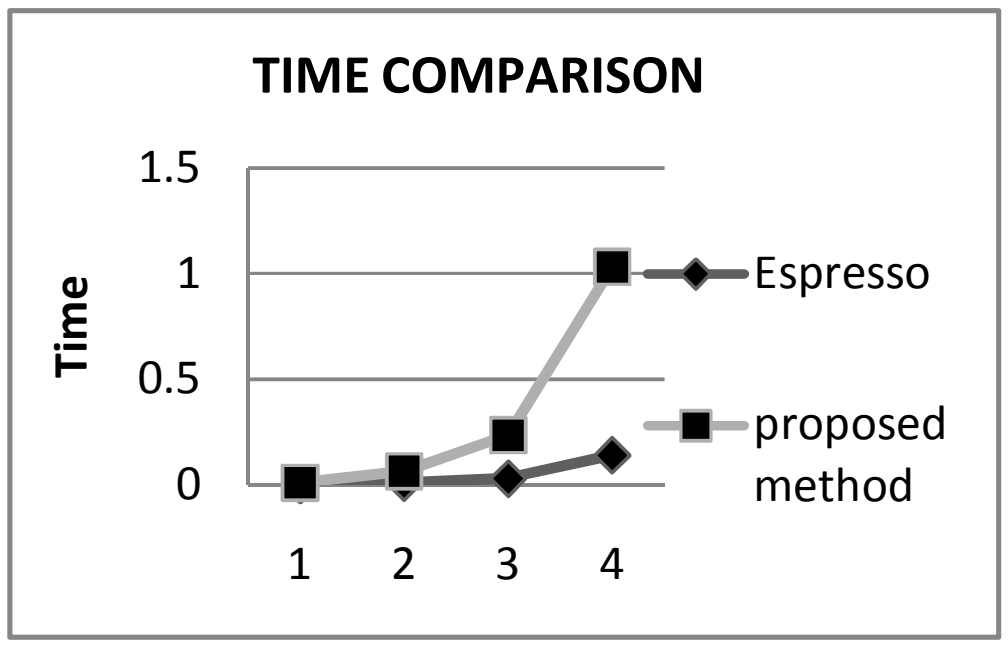

**(a)** **(b)**

**Fig 3:** **(a)** Comparison of number of disjoint cubes generated by ESPRESSO heuristic logic minimizer and the Proposed method. X-axis represents the number of variables and Y-axis represents the number of disjoint cubes. **(b)** Comparison of time taken to generate disjoint cubes by ESPRESSO heuristic logic minimizer and the Proposed method. X-axis represents the number of variables and Y-axis represents the time.

Here comparison of creation of number of disjoint cubes and time taken to create the disjoint cube of Espresso Logic Minimizer and the proposed method is done with respect to 4 variables.

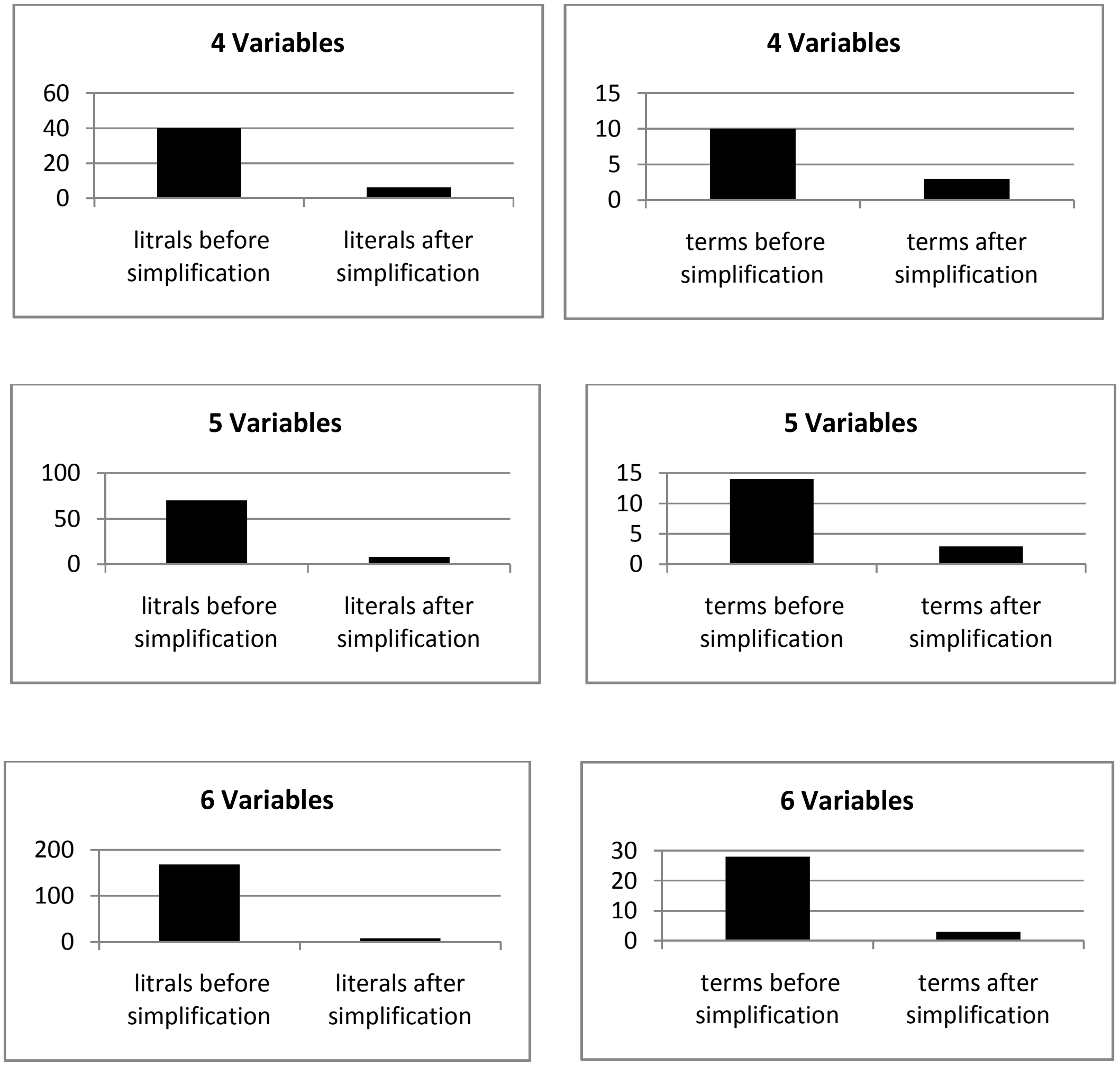

**Fig 4:** Comparison of the number of literals and terms before simplification and generated by above program for 4,5 and 6 variables.

Here number of literals and terms before and after minimizing the given Boolean sum-of-product function is given with respect to 4,5 and 6 variables with the Proposed method.

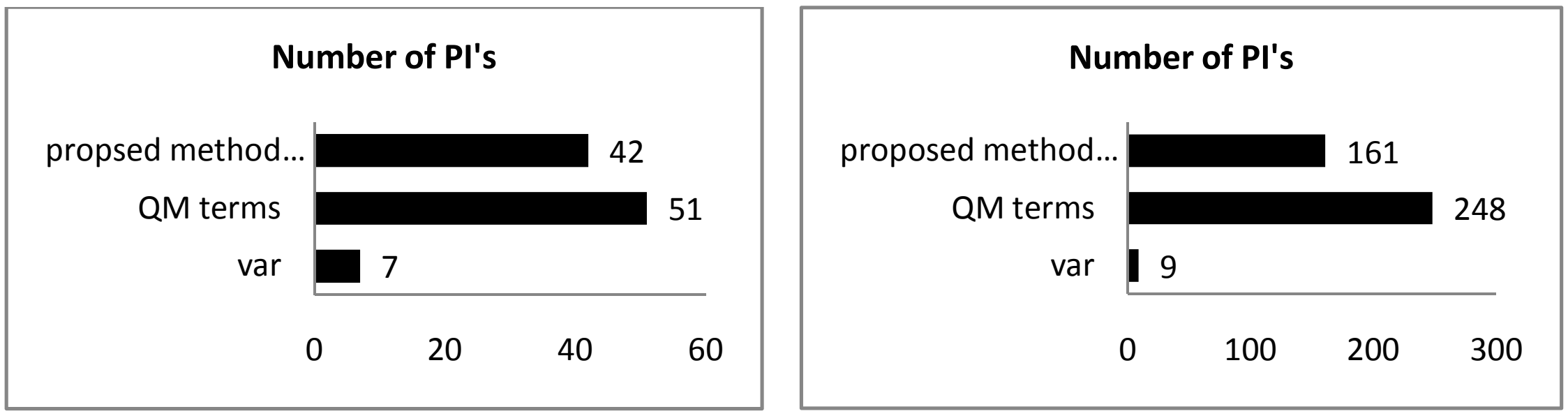

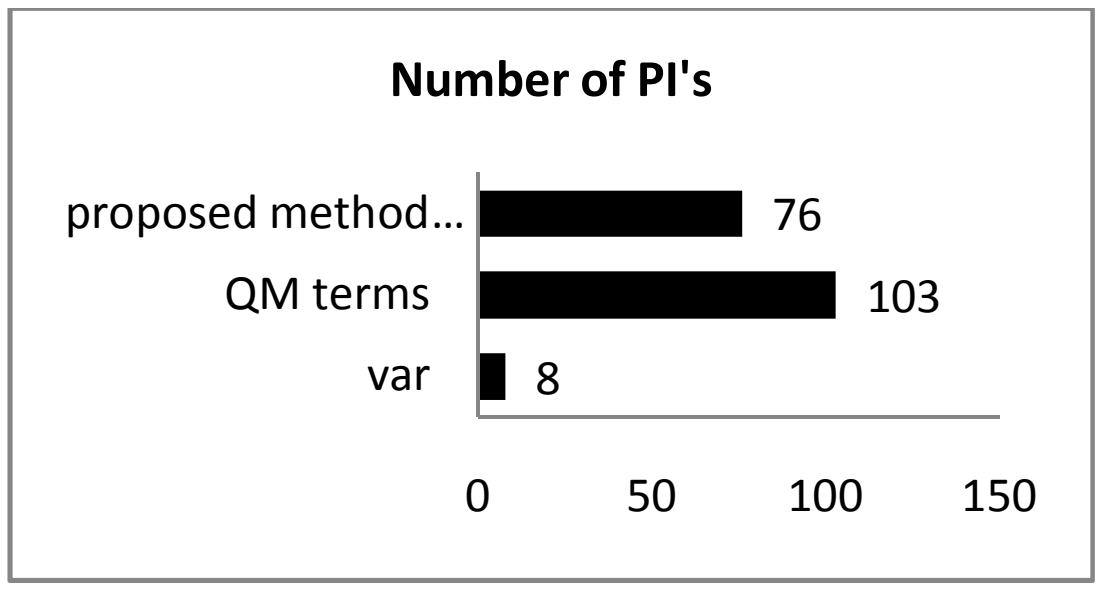

**Fig 5:** Comparison of number of Prime implicantes generated by Quine-McCluskey algorithm and the Proposed method for 7, 8 and 9 variables. Here QM stands for Quine-McCluskey.

Prime implicants generated by the Quine-McCluskey procedure and the Proposed method is compared here with respect to 7,8 and 9 variables.

## 6. Conclusion

An approach based on Binary Decision Diagram and Unate recursive paradigm with Binate covering algorithm to minimize the Boolean SOP function with DSOP representation of a Boolean function was presented. It is completely based on heuristics and gives the near optimum solution. But this procedure will only work for single output and completely specified functions and gives the near optimal solution. Whether it will work for incompletely specified function and multiple-output function is not tested yet. The work is in progress.

## 7. Future Works

1. Generation of all possible minimal covers or minimal expressions.
2. Compare with ESPRESSO logic minimizer.
3. Test whether it will work for incompletely specified function and multiple-output functions.

## Acknowledgements

The authors would like to thank West Bengal State University, West Bengal, India, Sarsuna College, West Bengal, India and University Of Calcutta, West Bengal, India. The authors also thank the reviewers for their constructive and helpful comments and specially the Computer without which no work was possible.

**Authors**

**Debajit Sensarma**
M. Sc. Computer Science

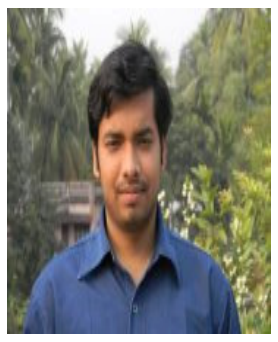

**Subhashis Banerjee**
M.Sc. Computer Science

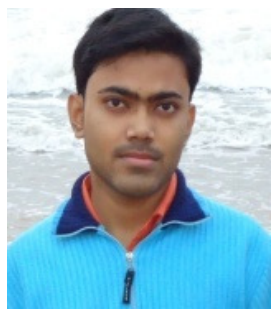

**Krishnendu Basuli**
Assistant Professor,
Department of Computer Science,
West Bengal State University.

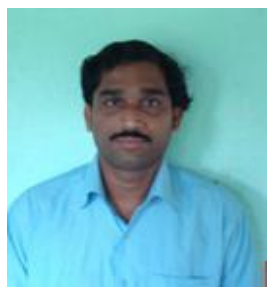

**Saptarshi Naskar**
Assistant Professor,
Department of Computer Science,
Sarsuna College, West Bengal, India.

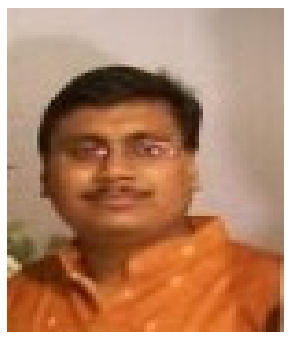

**Dr. Samar Sen Sarma**
Professor, Department of Computer Science and
Engineering, University Of Calcutta, India

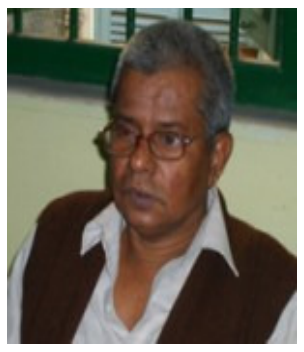